\documentclass[runningheads]{llncs}
\usepackage{graphicx}
\usepackage{amsmath}

\DeclareMathOperator*{\squareElemWise}{squareElemWise}
\DeclareMathOperator*{\softMax}{softMax}
\usepackage[all=normal,floats, charwidths, wordspacing]{savetrees}

\begin{document}
\title{Privacy-preserving Credit Scoring via Functional Encryption}
%
%
\author{
Lorenzo Andolfo\inst{1}\and
Luigi Coppolino\inst{1}\and
Salvatore D'Antonio \inst{1}\and 
Giovanni Mazzeo\inst{1}\and
Luigi Romano\inst{1} \and 
Matthew Ficke \inst{2} \and
Arne Hollum \inst{2} \and
Darshan Vaydia \inst{2} 
}
\authorrunning{L. Andolfo et al.}
%
\institute{
University of Naples 'Parthenope', Department of Engineering\\
\email{\{lorenzo.andolfo, luigi.coppolino, salvatore.dantonio, giovanni.mazzeo, luigi.romano\}@uniparthenope.it}\\
\and
X-Margin Inc., 141 Nantucket Cove, San Rafael, California, 94901\\
\email{\{matthew, arne, darshan\}@xmargin.io}
}
\maketitle              
\begin{abstract}
The majority of financial organizations managing confidential data are aware of security threats and leverage widely accepted solutions (e.g., storage encryption, transport-level encryption, intrusion detection systems) to prevent or detect attacks. Yet these hardening measures do little to face even worse threats posed on \emph{data-in-use}. Solutions such as Homomorphic Encryption (HE) and hardware-assisted Trusted Execution Environment (TEE) are nowadays among the preferred approaches for mitigating this type of threats. However, given the high-performance overhead of HE, financial institutions ---whose processing rate requirements are stringent--- are more oriented towards TEE-based solutions. 
The \emph{X-Margin Inc.} company, for example, offers secure financial computations by combining the Intel SGX TEE technology and HE-based \emph{Zero-Knowledge Proofs}, which shield customers' \emph{data-in-use} even against \emph{malicious insiders}, i.e., users having privileged access to the system. Despite such a solution offers strong security guarantees, it is constrained by having to trust Intel and by the SGX hardware extension availability. In this paper, we evaluate a new frontier for \emph{X-Margin}, i.e., performing privacy-preserving credit risk scoring via an emerging cryptographic scheme: \emph{Functional Encryption (FE)}, which allows a user to only learn a function of the encrypted data. We describe how the \emph{X-Margin} application can benefit from this innovative approach and ---most importantly--- evaluate its performance impact.

\end{abstract}
\section{Introduction}
Data confidentiality has significant relevance in the financial field. The disclosure of sensitive information such as credit scoring or exchange apikeys, can lead to, e.g., learn about users' credit risk ---and therefore decide whether to lend money--- or ultimately money theft. A data breach inevitably leads to serious consequences affecting the reputation of the financial institution. 
The importance of cyber-security is even more crucial for the cryptocurrency industry, whose current valuation ---$1.8$\$ trillion in 2021--- attracts a rapidly increasing number of hackers. Only in the last two years, four large hacks occurred in the digital currency space\footnote{\url{https://www.investopedia.com/news/largest-cryptocurrency-hacks-so-far-year/}}.
\\Companies are therefore pushing on equipping their platforms with state-of-the-art security mechanisms. There are widely accepted solutions such as storage encryption, transport-level encryption, intrusion detection systems, firewalls, IP white-listing, which are used and help in preventing common cyber-attacks. However, there are still open issues ---especially in terms of \emph{data-in-use} protection against privileged attackers (e.g., a \emph{malicious insider})--- that requires more advanced approaches. Companies aim for methods where the customer does not need to trust them. There are technologies and cryptography schemes that can be leveraged for this purpose. In this regard, the adoption of hardware-assisted Trusted Execution Environments (TEE) (e.g., Intel SGX) \cite{sgx} and Homomorphic Encryption (HE) \cite{he} is making its way into the industrial community to ensure this kind of protection. HE is extremely powerful since it enables arbitrary computations on ciphered data and allows the user to trust no one. A TEE such as SGX provides also strong security guarantees enabling the execution of sensitive code in an isolated and measurable area of CPUs, with the only limitation that the user needs to trust the hardware manufacturers such as Intel and that the application’s state is sealed in the hardware \cite{7912672}. Unfortunately, the exclusive use of HE is not doable since it is affected by a non-negligible execution time overhead, and by a large cipher text expansion. Moreover, a major problem of HE is the so-called unverifiable conditional issue, which forces a program running on a third-party host, and processing homomorphically encrypted data, to request that a client decrypts intermediate functional results to proceed further in the execution \cite{vise}. This introduces additional synchronization points and performance overhead, and increases the risk of information disclosure (e.g. via side channel analysis of the program workflow). For this reason, the trend for the majority of financial companies is to prefer TEEs. 
\\The \emph{X-Margin Inc.}\footnote{\url{https://xmargin.io/}} company, instead, uses both security mechanisms. It offers secure credit risk scoring by shielding computations in Intel SGX enclaves and performs HE-based \emph{Zero-Knowledge Proofs} to offer cryptographic Proof of Security, Computation and Encrypted Input. In this paper, we explore an innovative approach for \emph{X-Margin} which would allow to replace the use of SGX, thus removing Intel from the chain-of-trust and being hardware agnostic:  \emph{Functional Encryption} (FE) \cite{fe}. This is a cryptographic scheme similar to HE that does both evaluation and decryption of the \emph{result of a function} at the same time, without leaking the private key for data decryption and without leaking information about the plaintexts. This means that ---unlike HE--- there is no need anymore of doing synchronizations between the client and the computation entity for intermediate results. 
In a FE model, there are special ``evaluation keys'' that only allow the functional evaluation of ciphered data. An additional advantage given by the FE scheme is that it enables \emph{Attribute-based encryption} (ABE) schemes, where the encrypted data is linked with a set of attributes and secret keys along with certain policies that allow to control which ciphertexts can be decrypted depending on the possessed attributes. Unlike the classical encryption systems, where ``\emph{access to the encrypted data is all or nothing, one can either decrypt and read the entire plain or one learns nothing at all about the plain other than its length}'', FE tries to change this paradigm allowing a more fine-grained access to encrypted data and improving its flexibility. 
We describe how FE can be leveraged for \emph{X-Margin} financial computations and ---most importantly--- evaluate the performance impact of this innovative scheme. We overview a solution that computes credit risk from an encrypted borrower's data with the FE, thus preserving data privacy in a cloud environment even from the company itself. As a first step, we implemented an ad-hoc neural network for the purpose of credit risk scoring, which acts as a binary classifier saying whether a loan defaults. The training is conducted on a plain dataset. The resulting weight configuration downstream the training step is saved for the second step. We built the functional encryption scheme using the weights configuration obtained during the first step. We used a symmetric private key quadratic multivariate polynomial scheme called SGP \cite{sga_reading_in_the_dark}. As from the definition of functional encryption, the result of decryption is not the record itself, but the evaluation of a function of it, that is represented by the credit risk associated to the borrower. At the end, we computed the credit risk via two FE functions: one that computes the score to say whether the borrower defaults, and the other one that computes the score whether the borrower doesn't default. Then, the final result is normalized among the scores in order to provide the probability of default. Results from experimental evaluation show that the FE entails a non-negligible overhead on the classification time. Using 20 attributes, the application took 17.3s to compute the score with 200 borrowers in the system. It is important to notice that the scoring classification job is highly parallelizable, thus performance can be improved using multiple nodes. 
\\The rest of this paper is organized as follows:
In Section \ref{background}, we provide a background on credit scoring and on the FE scheme. Afterwards, in Section \ref{xmargin}, the X-Margin case study is presented.  
In Section \ref{solution}, the architecture of the proposed solution is discussed. 
Section \ref{evaluation} presents results from the experimental evaluation.
Finally, Section \ref{conclusion} concludes the document.


\section{Background}\label{background}
In this section, we provide a background on the two main elements covered in this work, i.e., credit scoring and functional encryption. 

\subsection{Principles of Credit Scoring}
Credit scoring is a method that is historically used by banks to estimate the risk of lending money to an individual. It determines the ability of a person to repay the debts. The higher is the credit score, the higher is the probability of obtaining a loan with low interest rates. On the other hand, people with a lower credit score must pay higher interest rates on their loans. The process of giving a loan is influenced by many factors: characteristics of borrowers in terms of who they are, their economic situation, the amount of the requested loan, its purpose, and the collateral type. The risk is estimated using elements of quantitative and qualitative analysis. The quantitative analysis takes into account an assessment of the financial standing of customers. It could also consider cash flow analysis of  customers' accounts together with their credit history. While the qualitative assessment includes other information such as the education, employment, industry in which they operate, or the way of keeping accounts.

\subsection{Fundamentals of Functional Encryption}
Functional encryption (FE) is relatively a new encryption technique, which was introduced by Amit Sahai and Brent Water in 2008 \cite{FE_pres}. 
FE enables selective access control of sensitive data $d$ basing on specific functions $f(d)$. In an FE scheme, a decryption key $sfKey(f)$ is associated with the function $f$. Therefore, the decryption of an encrypted data $d$ through $sfKey(f)$, provides the function evaluation $f(d)$, and nothing more about $d$. 
Concerning differences between FE and FHE, from one side they result similar, because once the message has been encrypted, both evaluate a function of the message. However, FHE suffers from to the additional step of decrypting the evaluation of the function, since the result of its computation is still an encrypted result. With FE instead, the result of the decryption already represents the evaluation of the function. FE has been successfully applied to cloud environments and to obfuscation mechanisms \cite{obfuscation}, that can be useful for example to protect intellectual property. Unfortunately, up to now, industries seldom adopt such advanced cryptography technologies: the majority of cryptography technologies used by industries were developed in early 2000s \cite{Crypt_Eng}. 
Only recently, thanks to the ambitious project FENTEC (Functional ENcryption TEChnologies) \footnote{\url{https://fentec.eu/}}, there has been a tentative to propose FE solutions for privacy preserving in a wide range of sectors from clinical data \footnote{\url{https://github.com/fentec-project/Selective-Access-to-Clinical-Data}}, to public transportation \footnote{\url{ https://github.com/fentec-project/FE-anonymous-heatmap}}.
\\There are several types of schemes that can be assimilated to the FE, these are: attribute-based encryption (ABE) \cite{ABE_M_L_MAPS}, \cite{FAME}, \cite{attr_based_encr_fine_grained_access}, identity-based encryption (IBE) \cite{IBE}, the ones that implement inner product functions \cite{simple_func_encr_schema_inner_prod}, \cite{multi_input_func_encr_inner_prod} and nonlinear (at most quadratic) polynomials. In this work, the focus is on a symmetric private key FE scheme called SGP, that implements a two dimensional quadratic polynomial function. 
\begin{equation} \label{eq:1}
	f(x,y): \sum_{i=1,j=1}^{m,n} f_{i,j} x_i y_j = \underset{1\times m}{X^T} \cdot \underset{m\times n}{\mathrm{F}} \cdot  \underset{n\times 1}{Y}
\end{equation}
Here, $X$ and $Y$ represents the vectors we want to encrypt, while $F$ is a custom matrix. The result of decryption in this case will be the evaluation of the function above.

Regarding security, it provides an adaptive security under chosen-plaintext attacks (IND-CPA security) which is based on bilinear pairings, with the following primitives:

\begin{enumerate}
\item   $msk \leftarrow GenerateMasterKey()$: generates a secret key for the SGP scheme.
    \item $c \leftarrow Encrypt(X, Y, msk) $: encrypts input vectors $X$ and $Y$ with the secret key $msk$ and returns the appropriate ciphertext $c$.
    \item 	$feKey \leftarrow DeriveKey(msk, F) $: given in input a custom matrix $F$ and the secret key $msk$, derives the FE key $feKey$.
    \item $ X^T \cdot \mathrm{F} \cdot Y \leftarrow Decrypt(c, feKey, F) $: given in input the ciphertext $c$ the FE key $feKey$ and the custom matrix $F$, gets the evaluation of $f(x,y)$ that is $ X^T \cdot \mathrm{F} \cdot Y$.
\end{enumerate}

\section{The X-Margin Case Study}
\label{xmargin}
\begin{figure}[hbt!]
\centering
\scalebox{0.75}{\includegraphics[width=\textwidth]{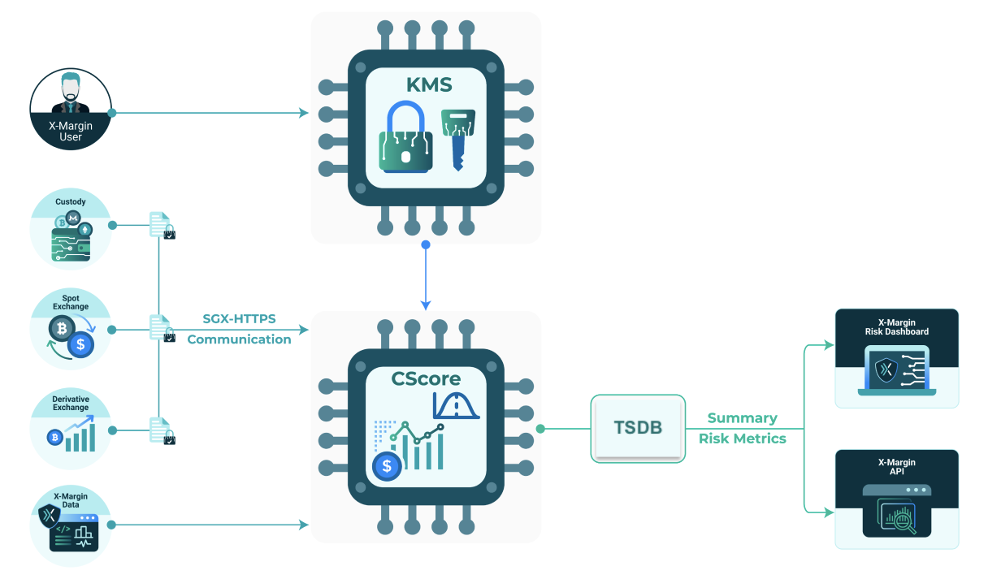}}
\caption{The \emph{X-Margin} architecture} \label{fig:xmargin}
\end{figure}

\emph{X-Margin Inc.} provides privacy-preserving credit scoring in the crypto-currency credit market. With over \$$100$B of crypto-collateral being used to generate over \$$1.25$B of interest on a quarterly basis, credit is one of the most rapidly growing sectors of the emerging crypto-currency finance ecosystem. \emph{X-Margin} allows borrowers to supply lenders with real-time portfolio risk metrics, while preserving the privacy of trades, positions, and other sensitive information. Borrowers benefit from improved lending terms, as they can display their risk in real-time and assure lenders they are trading responsibly. Lenders benefit from increased visibility and real-time information. \emph{X-Margin} calculates a variety of Risk Metrics on each user's portfolio, including Equity, Balance, Margin Usage, Maximum Loss (SPAN or VaR calculations), Aggregate absolute Delta, and 
. The company uses Zero Knowledge technology to ensure that the credit scoring system is functioning in an unbiased and privacy preserving way, and as the users expect. 
\\\emph{X-Margin} leverages TEE technologies, along with HE-based cryptographic Proof of Security, Computation and Encrypted Input, to ensure the privacy of users' sensitive data and guarantee risk analysis. Intel Software Guard Extensions (SGX), is the enabling TEE technology that is currently used to protect and attest the sensitive computations. SGX is a recent extension of modern CPUs, which ensures confidentiality even against super-privileged users running as \emph{root}. 
\emph{X-Margin} uses HE-based schemes for Proof of Computation including a Proof of Encrypted Input. Together, the proofs demonstrate that a given function being executed within the SGX enclave corresponds exactly to a symbolic representation stored in the server, outside the enclave ---the untrusted 'host'--- and that the data on which the function operates is received encrypted from an external source. \emph{X-Margin} proves that the functions the system is instructed to run are run in the enclave, without revealing the functions themselves. The Proof of Computation \& Encrypt
ed Input uses the Enhanced Skyline protocol, which is a proprietary protocol developed by \emph{X-Margin}'s team of cryptographers. Figure \ref{fig:xmargin} shows the current architecture of the \emph{X-Margin} solution. There is a SGX-based Key Management System (KMS) that holds users' apikeys, and a CScore unit where the actual private computation takes place, protected by the SGX secure enclave, and proven by cryptographic proofs. It queries spot/derivative exchanges and provide risk score metrics to a Time Series Database (TSDB). 
\\The goal of the company is to further improve their solution by raising the concept of \emph{trust} to a higher level. In the current situation, in fact, its customers must trust the hardware manufacturer, i.e., \emph{Intel}. It would be ideal for the company to remove also Intel from the \emph{chain-of-trust}. Furthermore, in the current status, the credit scoring application is strictly bounded to a particular hardware where the SGX ISA extension ---and the related motherboard support--- is available. This also entails migration constraints because the application's state is sealed in the hardware. \emph{X-Margin}'s R\&D is therefore exploring new innovative security techniques to cope with the mentioned limitations.

\section{The FE-based Credit Scoring Solution}
\label{solution}
The functional encryption paradigm represents an opportunity to overcome limitations posed by the SGX-based credit scoring solution. It ensures security properties similar to HE but with the advantage that it allows to learn specific functions of the encrypted data. Overall, the FE-based credit scoring works as follows. In a first phase, the neural network is trained leveraging plain datasets in \emph{X-Margin}'s premises. This is acceptable since the training process does not use any customers' sensitive data but it is conducted on anonymous data. At the end of this learning phase, coefficients needed to compute the credit risk of a given borrower are obtained. These are then used to generate keys. In this regard, according to the FE scheme principles, a trusted \emph{key authority} receives the coefficients and generates two types of keys, i.e.: public/private keys for the encryption/decryption of data, and operational keys needed to perform the functional computation on encrypted data, which can only be used to decrypt the result of the computation but not the data itself. At the end of this phase, the \emph{X-Margin} credit scoring application is finally configured. From now on, the customer will  encrypt the data along with the public keys received by the authority. The ciphered information is then sent to the untrusted \emph{X-Margin} premises running on cloud. Here, the credit risk associated to the borrower is computed on ciphered data. It  is important to notice that the computation of the credit risk associated to the borrower does not include any confidential borrower's data. Only the result of this computation (i.e., the scoring evaluation) will be accessible to \emph{X-Margin}.

\subsection{Architecture}
\begin{figure}[hbt!]
\centering
\scalebox{0.35}{
\includegraphics{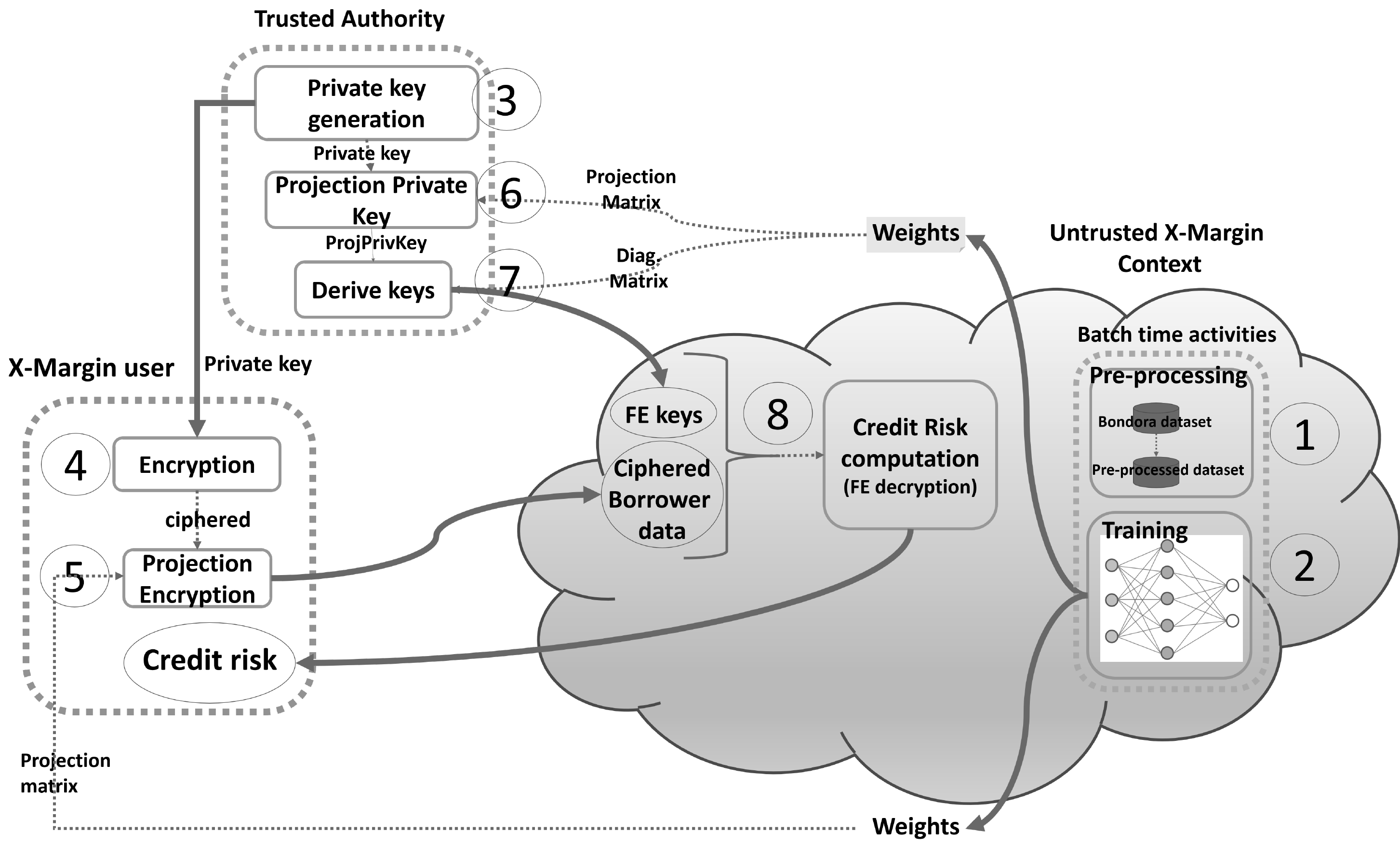}}
\caption{The credit risk score architecture} \label{fecscore}
\end{figure}

Figure \ref{fecscore} shows the overall architecture of the FE-based credit scoring solution, together with the steps performed during normal functioning. 

\begin{enumerate}
    \item \textbf{Data pre-processing}: At startup, \emph{X-Margin} performs a pre-processing activity on the data-set that will be used during the training phase.  
    Modifications to the dataset are needed to adapt the subsequent computation to the FE requirements, and also to improve the accuracy of the neural network model, such as:
    \begin{enumerate}
        \item Removes columns and rows containing null values, discard columns with a high percentage of null values, as well as rows with null values inside. 
        \item Removes outliers.
        \item Adapt timestamps
        \item Scales and normalizes columns.
        \item Replaces object columns with dummy variables and one-hot encodes the output.
        \item Transform data in integer form in order to ensure compatibility  with the FE scheme properties,
    \end{enumerate}

    \item \textbf{Neural Network Training}: In this phase, the training phase takes place using clear data. This is acceptable given that the leveraged dataset contains anonymous entries. An ad-hoc neural network was defined. We leveraged and adapted the \emph{Adam optimization algorithm}, with a learning rate of $10^{-4}$ and a batch size of $32$.   Training epochs have been set to $50$. The neural network has a single hidden layer and its activation layer is represented by a \emph{square element-wise}, instead of \emph{sigmoid}. This choice will be better explained in the rest of this section.
    \\Let $\underset{1 \times n}{X}$ be our input vector, that is the borrower's record, $\underset{n \times d}{Pr}$ be the weight matrix of the first layer and $\underset{d \times l}{D}$ be the weight matrix of the hidden layer. In our case, $n = 130$, $d = 20$ and $l = 2$. 
    The output of the neural network is:
    \begin{align} \label{eq:2}
	    prediction: \squareElemWise(\underset{1 \times n}{X} \cdot \
	    \underset{n\times d}{Pr}) \cdot \underset{d\times l}{D}
	    & = \underset{1 \times l}{DefaultScores}\nonumber.
    \end{align}
    The result of the classification is represented by a pair of scores, since $l = 2$, which are mutually exclusive and that are not normalized. In order to get the final result in terms of probability, a softmax function has been applied in the untrusted context side.

    At the end of the training step, $Pr$ and $D$ are saved and used afterwards for the encryption task and for deriving FE keys in the trusted context.
    \item \textbf{Secret Key Generation} The trusted authority generates a secret key for the symmetric SGP scheme according to the FE scheme, that is: 
      \begin{align}
      msk \leftarrow GenerateMasterKey()
      \end{align}
    It then sends $msk$ to the \emph{X-Margin} user. 
    \item \textbf{Data Encryption} The X-margin user uses $msk$ to encrypt the borrower's data $X$ and returns the appropriate ciphertext that is:  
    \begin{align}
      c \leftarrow Encrypt(X, X, msk)
      \end{align}
    Here, respect to the classical SGP scheme, we have replaced the parameter $Y$ by $X$. 
    \item \textbf{Encryption Projection} The \emph{X-Margin} user takes the encryption $c$ and weight matrix $Pr$ obtained at step 2 to produce a projection of $c$, that is: 
     \begin{align}
      ProjC \leftarrow projectEncryption(c, Pr)
      \end{align}
    $ProjC$ represents the encryption of $(Pr \cdot X)$, that is sent to the untrusted context.
    \item \textbf{Secret Key Projection} The trusted authority takes the secret key $msk$ obtained at step 3 and the weight matrix $Pr$ obtained at step 2 to get a projection of the secret key: 
    \begin{align}
      ProjSecKey \leftarrow projectSecKey(msk, Pr)
      \end{align}
    $ProjSecKey$ represents the secret key for encryption of $(Pr \cdot X)$.
    \item \textbf{FE Key Generation} The trusted authority provides also the FE keys to the untrusted party to compute the credit risk of the borrower. There are two FE keys: one used to decrypt the score of default and one used to decrypt the score of not default of the borrower. Therefore, each FE key is used to decrypt a different output $i$ of the neural network:
      \begin{align}
      feKey_i \leftarrow DeriveKey(ProjSecKey, Diag_i) \forall i \in l
      \end{align}
    $DeriveKey$ accepts as parameter the secret key $ProjSecKey$, obtained at the previous step, and the diagonal matrix $Diag_i$, that is the diagonalized version of $D_i$, that represents the $i-th$ row of the matrix $D$. It should be noted that $D_i$ is the vector containing the weights connected to a single label.
    
    Then, the trusted authority sends the obtained FE keys to the untrusted context.  
    \item \textbf{Credit Score Computation} The untrusted context, computes the credit risk scores of the borrower through the FE key provided by the trusted authority at the previous step. This means that for each label $i$ it evaluates the function of $(Pr \cdot X)$, that is $(Pr \cdot X)^T  \cdot Diag_i \cdot (Pr \cdot X)$. In order to do that, it exploits the the following decryption primitive of the SGP scheme:
    \begin{align} \label{eq:3}
        defaultScore_i & = (Pr \cdot X)^T  \cdot Diag_i \cdot (Pr \cdot X)\\        
                       & \leftarrow Decrypt(ProjC, feKey_i, Diag_i)\nonumber \forall i \in l\nonumber
    \end{align}
    For each label $i$, $Decrypt$ accepts $ProjC$, that as we saw represents the encryption of $(Pr \cdot X)$ sent by the \emph{X-Margin} user, the derived FE key $feKey_i$ and the diagonal matrix $Diag_i$ sent by the trusted authority. 

    Since $l = 2$, there are two functions: one that computes the score that the borrower defaults and one that he/she doesn't default. Such scores can be seen as mutually exclusive, therefore we can apply the soft max function on them in order to normalize the results and obtain the related probabilities. 

    \begin{equation}\label{eq:4}
        \underset{1 \times l}{DefaultProbabilities} = \softMax(\underset{1 \times l}{Scores}) 
    \end{equation}
    Finally, the untrusted context sends back to the \emph{X-Margin} user the score obtained at \ref{eq:4}. 
\end{enumerate}

\subsection{Details on the FE Scheme for Credit Risk Scoring}
The aim of this section is to draw connections between the neural network output function provided at \ref{eq:2} and the FE function at \ref{eq:1}, so as to understand what is the information needed to build the FE scheme. 

Given the FE function at \ref{eq:1}, we can transform it from a multivariate to a single variable function.
\begin{equation} \label{eq:5}
	f(k): \underset{1\times d}{K^T} \cdot \underset{d\times d}{\mathrm{F}} \cdot  \underset{d\times 1}{K}
\end{equation}
From the other side, we can see the neural network function at \ref{eq:2} as the the result obtained for each single label:
\begin{equation} \label{eq:6}
	defaultScore_i: \squareElemWise(\underset{1 \times n}{X} \cdot \underset{n\times d}{Pr}) \cdot \underset{d\times 1}{D_i} \forall i \in l.
\end{equation}
Where $D_i$ is a vector containing the weights connected to a single label $i$, $X$ is the borrower's vector and $Pr$ is the weight matrix associated to the first layer.
Applying the following manipulation to each function $i$ obtained in \ref{eq:6}, it is possible to get the FE function at \ref{eq:5}:

\begin{align} \label{eq:7}
	defaultScore_i & : \squareElemWise({X} \cdot {Pr}) \cdot {D_i}\\
	           & = ((Pr)^T \cdot (X)^T)^T \cdot {Diag_i} \cdot ((Pr)^T \cdot (X)^T)\nonumber\\ 
	           & = K^T \cdot Diag_i \cdot K\nonumber\\ 
	           & \forall i \in l\nonumber.
\end{align}
Here, we have transformed the first matrix product, switching the operands and applying the transpose on them. Indeed, we have removed the square element-wise operation on it and, in order to keep the equality we have replaced $D_i$ with its diagonalized version $Diag_i$ and we have multiplied by the product $((Pr)^T \cdot (X)^T)$. Finally, we have replaced the products with the obtained vector $K$ in order to highlight the equivalence with the \ref{eq:5}. The custom matrix $F$ is the diagonal matrix $Diag_i$ that is different for each label, while the other parts of the function don't change.  

We have shown that such neural network architecture computes the FE function \ref{eq:5}, for each label $l$. Each function has, as variable, the matrix product $((Pr)^T \cdot (X)^T)$ named $K$, and produces as output respectively the score that a borrower defaults and not. From now on $((Pr)^T \cdot (X)^T)$ will be indicated without the transpose operations inside, in order to not burden the discussion.

\section{Experimental Performance Evaluation} \label{evaluation}
We conducted an experimental evaluation of the privacy-preserving scoring solution. The goal was to estimate the performance overhead given by the adoption of the FE-based processing. More precisely, our focus was on the execution time needed to encrypt borrowers' data, and to compute the credit risk score, i.e., the evaluation of the functional decryption occurring on \emph{X-Margin} premises. These tests did not take into account the classification algorithm accuracy since our main goal is to focus on the sole performance. 

\begin{figure}[hbt!]
    \centering
    \scalebox{0.8}{
    {\includegraphics[width=\textwidth]{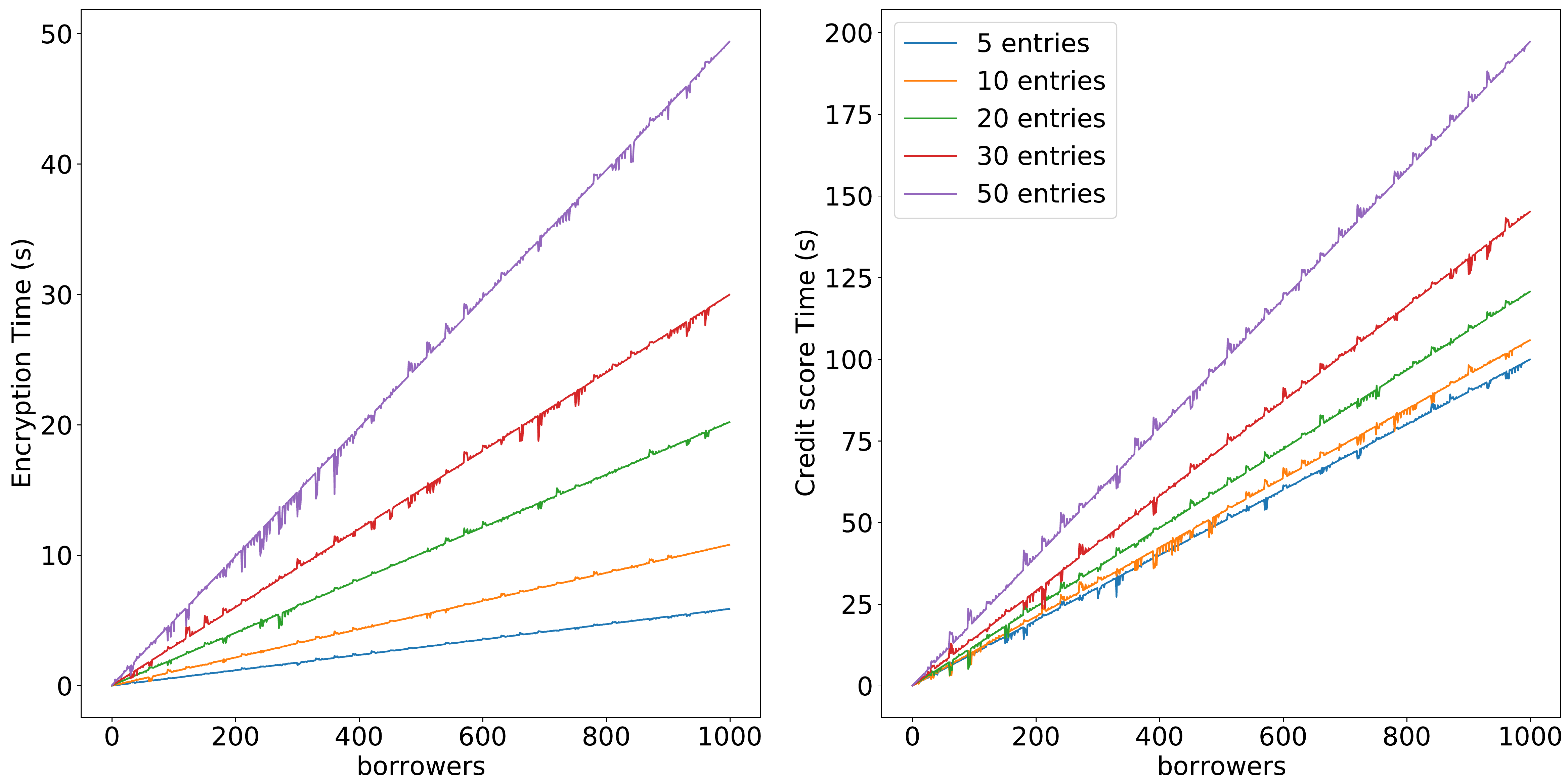}}}
    \caption{Encryption and credit scoring performance} \label{fig:enc_score}
    \label{fig:eval}
\end{figure}
We have evaluated our model on a Windows 10 machine equipped with an AMD Ryzen 3600X with 6-Core Processor at 3.80 GHz and 16GB of RAM. Execution times (in ms) are the average of 5 repeated tests. 
\\As a first experiment, we evaluated the encryption phase (i.e., steps 4 \& 5) by increasing the number of users in the system up to 1000 borrowers, and also varying the number of borrowers' attributes (in the range [5,25]). 
\\Figure \ref{fig:eval} shows the outcomes of the evaluation. It can be noticed that the encryption is not significantly time consuming. The encryption of one borrower in what we defined the worst situation ---i.e., characterized by 50 attributes--- takes on average $52.3ms$, while in the best situation ---i.e., 5 attributes--- it takes $14.3ms$. The execution time increases linearly with the number of borrowers. 
\\The second evaluation we made refers to the FE-based scoring computation time, which is critical for the actual usability of the proposed solution. It is important to notice that this operation includes both the actual functional computation and the final result decryption. Figure \ref{fig:eval} shows that the overhead is significant, reaching in the worst condition ---i.e., 50 attributes and 1000 users--- a computation time of $~170s$. It must be noticed that in real situations, the number of attributes is lower, in the order of 20. Nevertheless, performance are still not acceptable. However, the type of job we implemented with FE is highly parallelizable, and the deployment on multiple nodes can make the adoption of the FE-based credit scoring solution acceptable.

\section{Conclusion}
\label{conclusion}
In this paper, we presented an innovative solution for privacy-preserving credit risk scoring, which leverages the emerging \emph{Functional Encryption} cryptography that allows to learn the result of a specific function using only encrypted data. We built a credit risk model using the quadratic polynomial FE symmetric scheme SGP for the X-Margin Inc. company, whose case study was used to validate the effectiveness of the proposed approach. Results from the experimental campaign show that the solution performs well under certain conditions, but the overhead needs to be properly managed in case of a real production deployment. We discovered that with a limited number of borrowers' attributes (e.g. $<20$) the encryption time and the classification time are acceptable.  Future developments of the current solution will be focused on improving performance by evaluating new public key quadratic schemes (e.g., \cite{new_paradigm_publ_key_Func_degree_2}), and setting up a multi-node parallel architecture.

\bibliographystyle{splncs04}
\bibliography{references}

\end{document}